\documentclass{article}
\usepackage{times}
\usepackage{helvet}
\usepackage{courier}
\usepackage{graphicx}
\usepackage{caption}
\usepackage{color, indentfirst,xspace}
\usepackage{amsmath}
\usepackage{amsfonts}
\usepackage{amssymb}
\usepackage{calligra}
\usepackage{calrsfs}
\usepackage{algpseudocode}
\usepackage{algorithm}
\usepackage{subfig}
\usepackage{bbm}
\usepackage{verbatim}
\usepackage{euscript}
\usepackage{soul}
\usepackage[normalem]{ulem}
\usepackage{todonotes}

\makeatletter
\def\eytanf{\bgroup \markoverwith{\lower3.5\p@\hbox{\sixly \textcolor{red}{\char58}}}\ULon}
\font\sixly=lasy6 
\makeatother

\newcommand\redout{\bgroup\markoverwith
{\textcolor{red}{\rule[.5ex]{2pt}{0.4pt}}}\ULon}

\definecolor{purple}{RGB}{188, 19, 254}

\newcommand{\etal}{{\em et al.}}
\newcommand{\naive}{na\"{i}ve}

\newcommand{\eg}{{\em e.g.}}

\newcommand{\system}{\textsc{EdgeBoost}\xspace}
\let\mathcal = \EuScript

\begin{document}

\title{ Link-Prediction Enhanced Consensus Clustering for Complex Networks}

\author{
Matthew Burgess, Eytan Adar, Michael Cafarella\\
 University of Michigan, Ann Arbor\\
 \{mattburg, eadar, michjc\} @umich.edu
}

\maketitle
\begin{abstract}
Many real networks that are inferred or collected from data are incomplete due to missing edges. Missing edges can be inherent to the dataset (Facebook friend links will never be complete) or the result of sampling (one may only have access to a portion of the data).  The consequence is that downstream analyses that consume the network will often yield less accurate results than if the edges were complete. Community detection algorithms, in particular, often suffer when critical intra-community edges are missing. We propose a novel consensus clustering algorithm to enhance community detection on incomplete networks. Our framework utilizes existing community detection algorithms that process networks imputed by our link prediction based algorithm. The framework then merges their multiple outputs into a final consensus output. On average our method boosts performance of existing algorithms by 7\% on artificial data and 17\% on ego networks collected from Facebook.   

\end{abstract}
\section{Introduction}

Many types of complex networks exhibit community structures: groups of highly connected nodes. Communities or clusters often reflect nodes that share similar characteristics or functions. For instance, communities in social networks can reveal user's shared political ideology~\cite{Conover_Political}. In the case of protein interaction networks, communities can represent groups of proteins that have similar functionality~\cite{Jonsson_Cluster}. Since networks that exhibit community structure are common in many disciplines, the last decade has seen a profusion of methods for automatically inferring community structure.

Community detection algorithms rely on the topology of the input network to identify meaningful groups of nodes. Unfortunately, real networks are often incomplete and suffer from missing edges. For example, social network users seldom link to their complete set of friends; authors of academic papers are limited in both the number of papers they can cite, and can clearly only cite already-published papers. Missing edges can also be a result of the data collection process. For instance, Twitter often limits its data feed to only a 10\% ``gardenhose'' sample: constructing the mention graph from this data would yield a graph with many missing edges~\cite{Morstatter_Is}. Datasets crawled from social networks with privacy constraints can also lead to missing edges. In the case of protein-protein interaction networks, missing edges result from the noisy experimental process used to measure pairwise interactions of proteins~\cite{Huang_Where}. Community detection algorithms rarely consider missing edges and so even a ``perfect'' detection algorithm may yield wrong results when it infers communities based on incomplete network information.

One straightforward approach for improving community detection in incomplete networks is to first ``repair'' the network with {\em link prediction}, and then apply a community detection method to the repaired network~\cite{Mirshahvalad_Significant}.  The link prediction task is to infer ``missing'' edges that belong to the underlying true graph. A link prediction algorithm examines the incomplete version of the graph and predicts the missing edges. Although link prediction is a well-studied area~\cite{Lu_Link,Nowell_Link}, little attention has been given to how it can be used to enhance community detection. Imputing missing edges using link prediction can result in the addition of both intra- and inter-community links. If one were to simply run a link predictor and cluster the resulting network, the output can only be improved if the link predictor accurately predicts links that reinforce the true community structure.

\begin{figure}[t]
\centering
\includegraphics[width=\columnwidth]{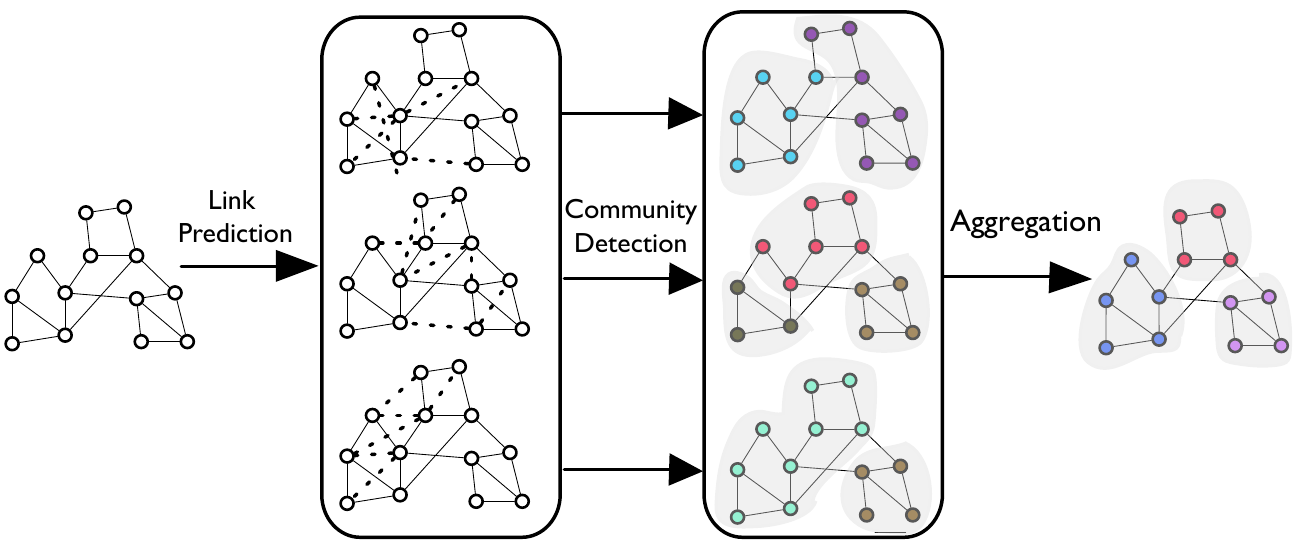}
\caption{Diagram describing work flow of \system}
\label{fig:system_diagram}
\end{figure}

We propose the \system\ method, which repeatedly applies link prediction via a random process, thereby mitigating the inaccuracies in any single link-predictor run. Our method first uses link prediction algorithms to construct a probability distribution over candidate inferred edges, then creates a set of imputed networks by sampling from the constructed distribution. It then applies a community detection algorithm to each imputed network, thereby constructing a set of community partitions. Finally, our technique aggregates the partitions to create a final high-quality community set   

An important and desirable quality of our method is that it is a {\it meta-algorithm} that does not dictate the choice of specific link prediction or community detection algorithms.  Moreover, the user does not have to manually specify any parameters for the algorithm. We propose an easy-to-implement, black-box mechanism that attempts to improve the accuracy of any user-specified community detection algorithm. Our contributions in this paper include:

\section{Related Work}
\label{sec:related_work}

{\bf Community Detection Overview} --- There are many variants of the community detection problem: communities can be disjoint, overlapping, or hierarchical. The problem of detecting disjoint communities of nodes is the most popular and what we focus on in this work. While the other variants, especially overlapping community detection, are of growing interest, detecting strict partitions is still a hard and relevant problem. In fact, recent work~\cite{Prat_Perez_High} has shown that disjoint algorithms can perform better than overlapping algorithms on networks with overlapping ground truth. We chose a collection of six algorithms to test our system on: Louvain~\cite{Blondel_Fast}, InfoMap~\cite{Rosvall_Mapping}, Walk-Trap~\cite{Pons_Computing}, Label-Propagation~\cite{Raghavan_Near}, Significance~\cite{Traag_Significant} and Surprise~\cite{Aldecoa_Deciphering,Traag_Detecting}. We chose Louvain and Infomap for many of our experiments because both perform well in recent comparisons \cite{Aldecoa_Exploring,Lancichinetti_Community,Prat_Perez_High}; Infomap is typically  superior in quality while Louvain is more scalable.

{\bf Ensemble Community Detection} --- Though single-tech\-nique community detection is by far the most common, a number of recent projects have proposed ensemble techniques~\cite{Dahlin_Ensemble,Lancichinetti_Concensus,Aldecoa_Exploring}. Aldecoa \etal\ describe an ensemble of partitions generated by different community detection algorithms, which differs from our approach of using the same algorithm and creating the ensemble by creating different networks. Both~\cite{Dahlin_Ensemble} and~\cite{Lancichinetti_Concensus} present techniques for consolidating partitions generated by repeatedly running the same stochastic community detection algorithm. We implemented both of their methods but neither was suitable for consolidating clusters in our system; this is most likely because the partitions generated from our system have more variation than partitions generated from multiple runs of a stochastic algorithm. At a high-level, our proposed technique is a type of ensemble.  Most ensemble solutions take the network as-is and assume that a ``vote'' between algorithms will produce more correct clusters. While this may work in some situations, bad input will often reduce the performance of all constituent algorithms (possibly in a systematic way) and therefore the overall ensemble. Our proposed method is novel in its iterative application of link prediction to increase the efficacy of community detection algorithms.

{\bf Ensemble Clustering} --- Ensemble data clustering (for a survey see~\cite{Ghaemi_Survey}) first proposed by Strehl \etal~\cite{Strehl_Cluster} involves the consolidation of multiple partitions of the data into a final, hopefully higher quality partitioning. While many of the ensemble clustering methods share a similar work flow to our method, the fact that these techniques were developed for data clustering and not community detection make them distinct from our work. For instance, Dudoit \etal~use bootstrap samples of the data to generate an ensemble of partitions, which in the case of network community detection would be difficult since networks have an interdependency between nodes, and nodes cannot be sampled with replacement like data in euclidean spaces. Monti \etal~\cite{Monti_Concensus} propose a consensus clustering technique with the goal of determining the most stable partition over various parameter settings of the input algorithm. Similar to our work, many ensemble clustering algorithms ~\cite{Dudoit_Bagging,Fern_Solving,Monti_Concensus,Strehl_Cluster} use a consensus matrix as a data structure to aggregate the ensemble of partitions. Unlike previous methods ~\cite{Dudoit_Bagging,Strehl_Cluster}, which use agglomerative clustering to compute the final partition we propose an aggregation algorithm that uses connected components, which is not possible on data clustering problems.

{\bf Community Significance} --- In the community detection literature, techniques have been proposed for both evaluating the significance/robustness of communities as well for detecting significant communities. \cite{Karrer_Robustness} propose a network perturbation algorithm for evaluating the robustness of a given network partition. Methods have also been developed~\cite{Hu_measuring,Lancichinetti_Statistical} that measure the statistical significance of individual communities.  Our goal, however, is not to generate confidence metrics on communities but rather to generate more accurate communities overall.  
Previous work has also proposed techniques for finding significant communities using sampling based techniques~\cite{Gfeller_Finding,Mirshahvalad_Resampling,Rosvall_Mapping}. Rosvall \etal\ and Mirshahvalad \etal\ propose algorithms for detecting significant communities by clustering bootstrap sample networks and identifying communities that occur consistently amongst the sample networks. The method proposed by Gfeller \etal\ attempts to identify significant communities by finding unstable nodes using a method based on sampling edge weights. Their methods differ from ours in that they create samples from the {\it existing} network topology. Most similar to our work is the paper by~\cite{Mirshahvalad_Significant} which attempts to solve the problem of identifying communities in sparse networks by adding edges that complete triangles. Their method is simply to add a fixed percentage of triangle completing edges and cluster the resulting network; in contrast, our approach which involves the repeated application of any link prediction algorithm.

{\bf Community Granularity} --- The problem of detecting communities at various levels of granularity is a well studied and related problem. Work by ~\cite{Fortunato_Resolution,Xiang_Limitation} has analyzed the ``resolution limit'' of detecting communities at all granularities. As a result of these limitations many methods~\cite{Arenas_Analysis,Delvenne_Stability,Li_Quantitative,Ronhovde_Local} have been proposed for community detection at different granularities. New objective functions that improve the resolution limit~\cite{Li_Quantitative} as well as tunable objectives~\cite{Arenas_Analysis,Ronhovde_Local} that allow community detection at various resolutions have been proposed. Delvenne \etal\ propose a method for identifying the stability of communities by using the Markov time of a random walk on the network. Granularity is a related problem in that missing edges can lead to communities detected at wrong granularities. However, these methods do not address the problem of detecting communities on incomplete networks.

\section{Problem Formulation}
\label{sec:problem_formulation}
\subsection{Communities in Incomplete Networks}
\label{sec:baseline_experiment}

    To motivate the need for algorithms that are robust to missing edges, we experimented on existing community detection algorithms. To test these algorithm's sensitivity to missing edges on a range of networks, we utilize the LFR benchmark~\cite{Lanc_Benchmark}. LFR creates random networks with ground-truth community structure (planted partition), parametrized by:  number of nodes, mixing parameter $\mu$, and exponent of degree and community size distributions (see~\cite{Lanc_Benchmark} for a full description). The mixing parameter is a ratio that ranges from only intra-community edges (0) to only inter-community edges (1). Previous studies~\cite{Aldecoa_Exploring,Lancichinetti_Community} have compared the quality of community detection algorithms using the benchmarks and used $\mu$ as the variable parameter, roughly capturing how difficult a network is to cluster. As we are concerned with characterizing the effect of missing edges, we modify the LFR benchmark by randomly deleting edges from the networks it generates. We denote the parameter $\delta$ as the percentage of removed edges.

\begin{figure}[h*]
  \centering
  \includegraphics[scale = 0.48]{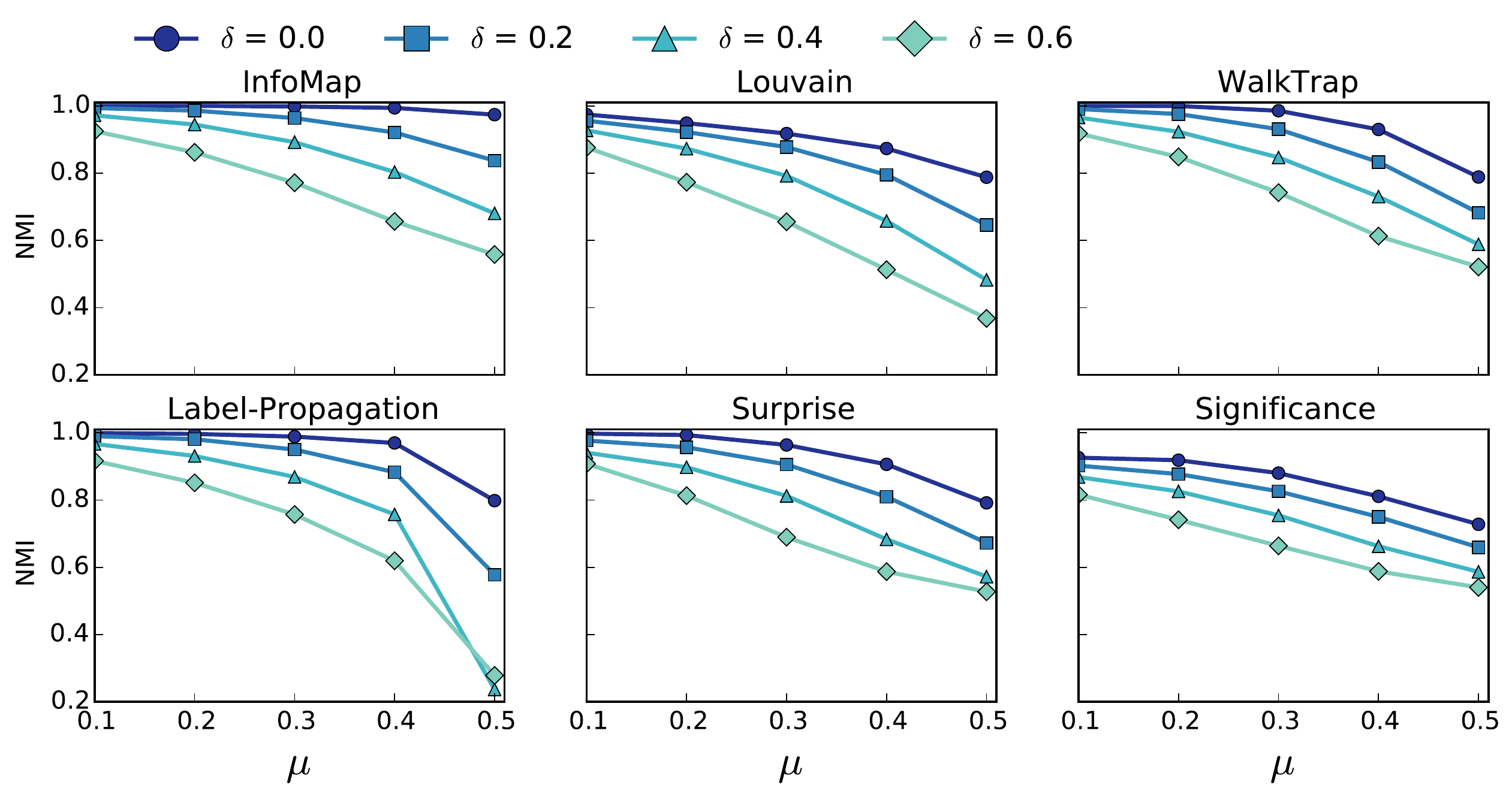}
  \caption{NMI of six community detection algorithms with varying percentages of removed edges.}
\label{fig:baseline_vs_noise_NMI}
\end{figure}

\begin{figure}[h*]
  \centering
  \includegraphics[scale = 0.48]{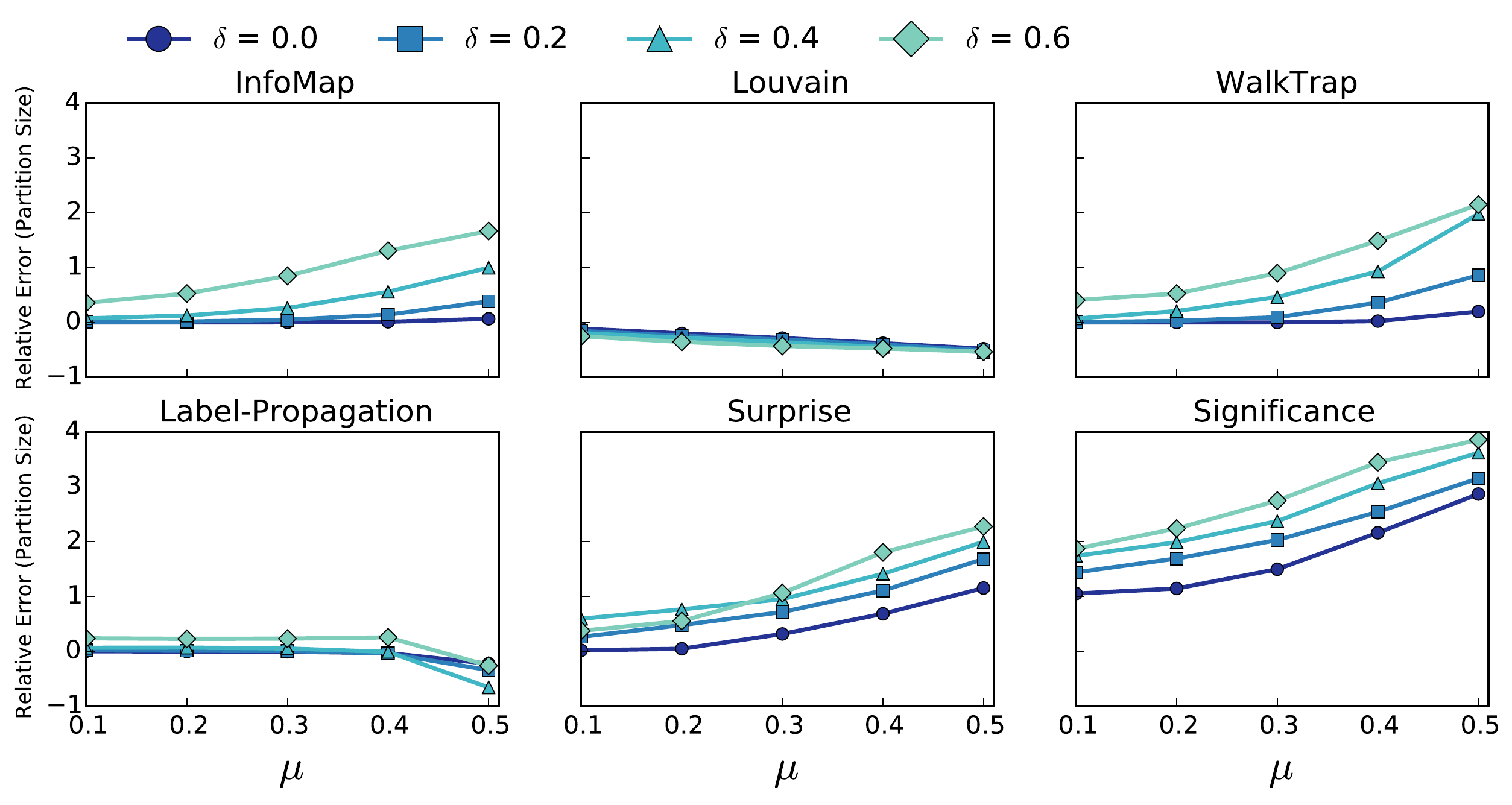}
  \caption{RE of six community detection algorithms with varying percentages of removed edges.}
\label{fig:baseline_vs_noise_RE}
\end{figure}
	
The goal of our analysis is to characterize the effect of both $\mu$ and $\delta$ on two metrics: Normalized Mutual Information (NMI) and the Relative Error (RE) of the size of the inferred partitions. NMI is a standard information theoretic measure for comparing the planted partition provided by the benchmark to the inferred partition produced by the algorithm. We define RE as the relative error of the number of communities inferred by the algorithm $C$ compared to the number of communities $C*$ in the planted partition:

\begin{equation}
 RE := \frac{C-C*}{C*}
\end{equation}

Since NMI can decrease for a variety of reasons (shifted nodes, shattered or merged communities), we include RE as a means to determine the more specific effects that missing edges can have on community detection. Each point in the plot is generated by averaging the corresponding statistic over 50 random networks generated by our modified LFR benchmark. We set static values for the following benchmark parameters: Number of nodes (1000), the average degree (10), the maximum degree (50), the exponent of the degree distribution (-2), exponent of the community size distribution (-1), minimum community size (10), and maximum community size (50). We varied parameters, such as ``number of nodes'' and ``average degree'', finding qualitatively similar results for the effect of $\delta$ on NMI and RE. Similar to previous research ~\cite{Mirshahvalad_Significant}, we select an ``average degree'' that results in the sparse networks that motivate the need for the methods presented in this paper.

Figure~\ref{fig:baseline_vs_noise_NMI} shows how NMI varies with respect to $\delta$ and $\mu$ for six popular community detection algorithms. All of the algorithms behave in a qualitatively similar manner, as $\delta$ increases, the NMI score decreases. Similar to previous studies, InfoMap scores best with respect to $\mu$ and not surprisingly is also the most robust to missing edges. More interesting are the results in Figure ~\ref{fig:baseline_vs_noise_RE} which show how the number of inferred communities differ with respect to the number of communities in the planted partition. Four of the six algorithms show a trend of detecting too many communities both as a function of $mu$ and $\delta$, while only the Louvain and Label-Propagation algorithms detect fewer communities on average. Modularity is known to suffer from a resolution limit \cite{Fortunato_Resolution}, meaning that the measure tends to favor larger communities. Since Louvain uses modularity as its objective function, it is not surprising that Louvain, on average, infers communities that are larger then in the planted partition. Overall, it is more often the case that missing edges will cause community detection algorithms to ``shatter'' ground truth communities, sometimes producing 2-3 times more communities. Both in terms of NMI and RE, all 6 algorithms show a significant deterioration in community detection quality, once again, underscoring the need for algorithms that are robust to missing edges.

\subsection{Link Prediction for Enhancing Community Detection}
\label{sec:link_prediction_quality}

The ideal scenario for community detection is one where a network consists of only intra-community edges and where the detection of communities reduces to the problem of identifying weakly connected components. The reality is that we rarely find such clean graphs as edges can be ``missing'' for anything ranging from sampling to semantics. This last factor is important as a missing edge between nodes in the same community is not necessarily incorrect--the semantics of a network does not necessitate an explicit relationship between users in the same community. In the case of an ego-network on Facebook, for example, not all friends in the same community actually know each other as they may be grouped because they attend the same college as the ego-user. Similarly, a biological network may have a set of proteins working in concert as part of a functional ``community'' but many do not form a clique as the edges represent (up or down)-regulation. In both scenarios The edges that are missing are implicit edges representing the intra-community links (e.g., an edge representing the relationship {\it in-the-same-community-as}). It is these intra-community edges, whether they are implicit or explicit, that can have severe impact on the detection of communities. 

The hypothesis of this paper is that by recovering edges in incomplete networks, community detection quality can be improved. If link prediction is to be an effective strategy at recovering lost community structure, it must be accurate at predicting intra-community edges that reinforce communities. If the link prediction algorithm has too high a false-positive rate, thereby predicting too many inter-community links, it is likely to degrade community detection performance. Using the modified LFR benchmark, we analyzed the intra-community precision of various link prediction algorithms over a range of $\mu$ and $\delta$ values. We do not intend to exhaustively test all of the link prediction algorithms proposed in the literature, but we select three computationally efficient techniques that are among the best \cite{Nowell_Link,Lu_Link}: Adamic-Adar (AA), Common-neighbors (CN), and Jaccard.

\begin{figure}[t]
  \centering
  \includegraphics[width=\columnwidth]{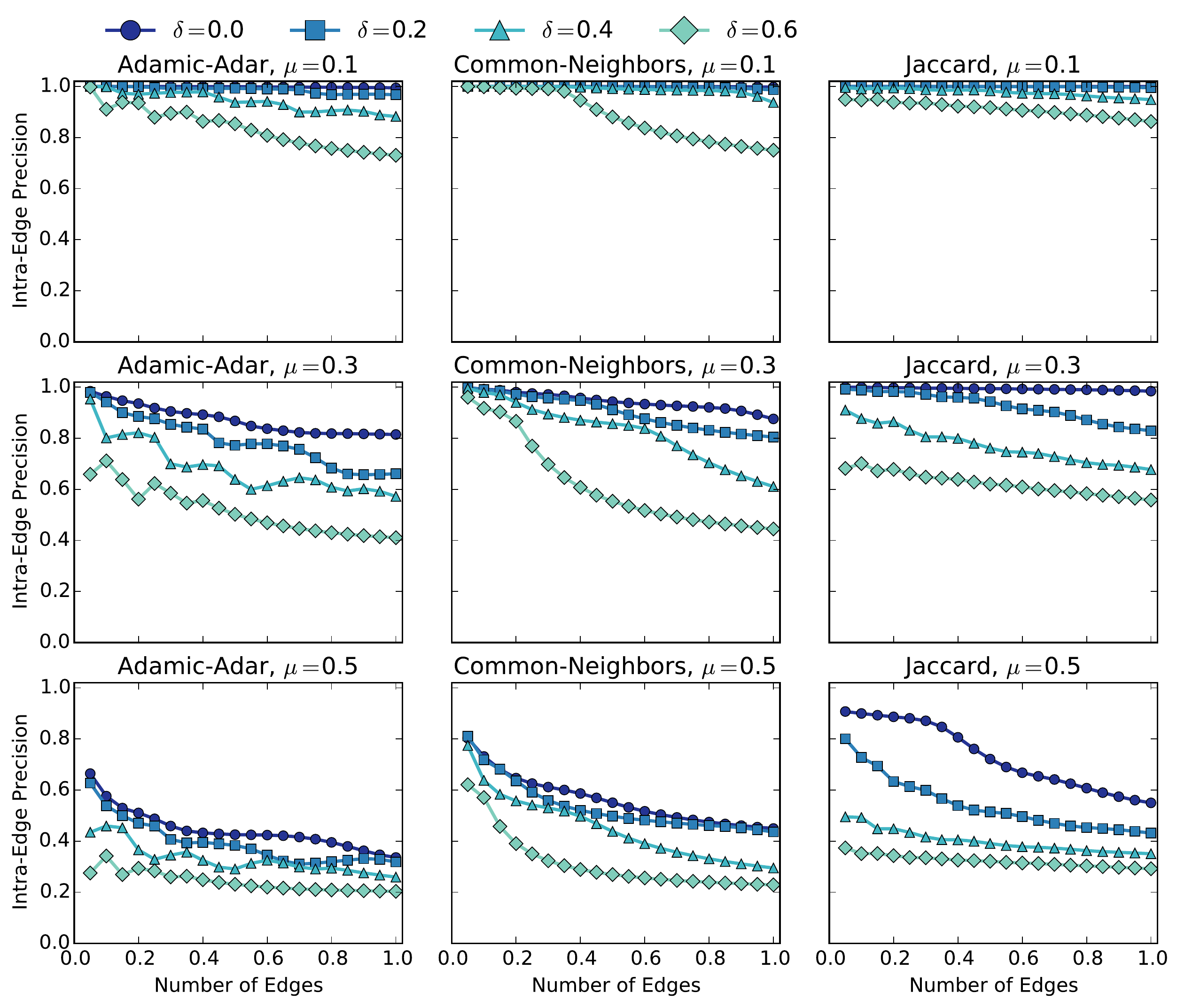}
  \caption{Precision plots of three link prediction algorithms: Adamic-Adar (left), Common Neighbors (middle), and Jaccard (Right) for various values of mixing parameter $\mu$: 0.1 (top), 0.3 (middle), and 0.5 (bottom). The $X$-axis corresponds to number of top-$k$ edges as scored by the link prediction algorithm as a percentage of the number of edges in the network. Intra-edge precision is on the $y$-axis.}
\label{fig:link_prediction}
\end{figure}

Each of these algorithms can produce a score for missing edges that complete triangles in the input network, allowing us to create a partial ordering over the set of missing edges. Figure~\ref{fig:link_prediction} shows the results from our experiment. For each plot, the $y$-axis represents the intra-edge precision and the $x$-axis represents the number of top $k$ edges as a percentage of the total number of edges in the original network (before random deletion). For example, if the original network had 2000 edges, then an edge percent value of 20\% would correspond to selecting the top-400 edges from the network as scored by the given link prediction method. By varying $k$ we are able to observe the classification accuracy inferred by the ranking produced by each link-predictor. 
	
In Figure~\ref{fig:link_prediction} we first notice that as with community detection, link prediction performance decreased as a function of both $\delta$ and $\mu$ value. For low $\mu$, all link prediction algorithms are capable of achieving high intra-edge precision even for $\delta$ values of 60\%, but the quality of link prediction drops significantly for high levels of $\mu$. For $\mu$  above 0.5, any link-predictor that uses the number of common-neighbors as a signal will do poorly, since the majority of a node's neighbors belong to different communities. The Jaccard algorithm maintains the highest level of precision as a function of the number edges. While the AA algorithm usually outperforms Jaccard, AA is only better for low values of $k$.

The results in Figure~\ref{fig:link_prediction} show that link prediction can be effective at imputing intra-community edges, especially for sparse networks that have lower $\mu$ values. The results also show that for networks with high $\mu$ and $\delta$ values, the top-scoring edges as predicted by all three link prediction algorithms contain a large percentage of inter-community links. While this demonstrates the feasibility of using link prediction to recover missing intra-community edges we do not know how to set the parameters (e.g., the $k$ value to use for partitioning the ranked edges) for real-world networks. We will return to this, but first we formalize the problem.

Let $G = (V,E)$ be the input network, and the set $E_\text{missing} = (V \times V) \setminus E$ denote the set of missing edges in the $G$. We formally define a link-predictor $\EuScript{L}$ as a function that takes any pair of nodes in $(x,y) \in E \setminus V \times V$ and maps them to a real number.
\begin{equation}
 \mathcal{L}: E_\text{missing} \mapsto \mathbb{R}
\end{equation}
A community detection algorithm can be formally described as a function $\mathcal{C}$ that takes as input any network $G$ and produces a disjoint partition of the nodes $\{C_1,C_2,...,C_k\}$.

The most \naive\ algorithm for enhancing community detection consists of a few simple steps. First, score missing edges in $G$ using $\mathcal{L}$. Next, select the top-$k$ missing edges according to the link-predictor and add these edges to $G$. Lastly, apply the algorithm $\mathcal{C}$ to the imputed network.  However, simply adding links with high scores for networks with large $\mu$ and $\delta$ values may be problematic, since many of these links can be inter-community, thereby having a negative effect on community detection.

\begin{figure}[t]
  \centering
  \includegraphics[scale=0.35]{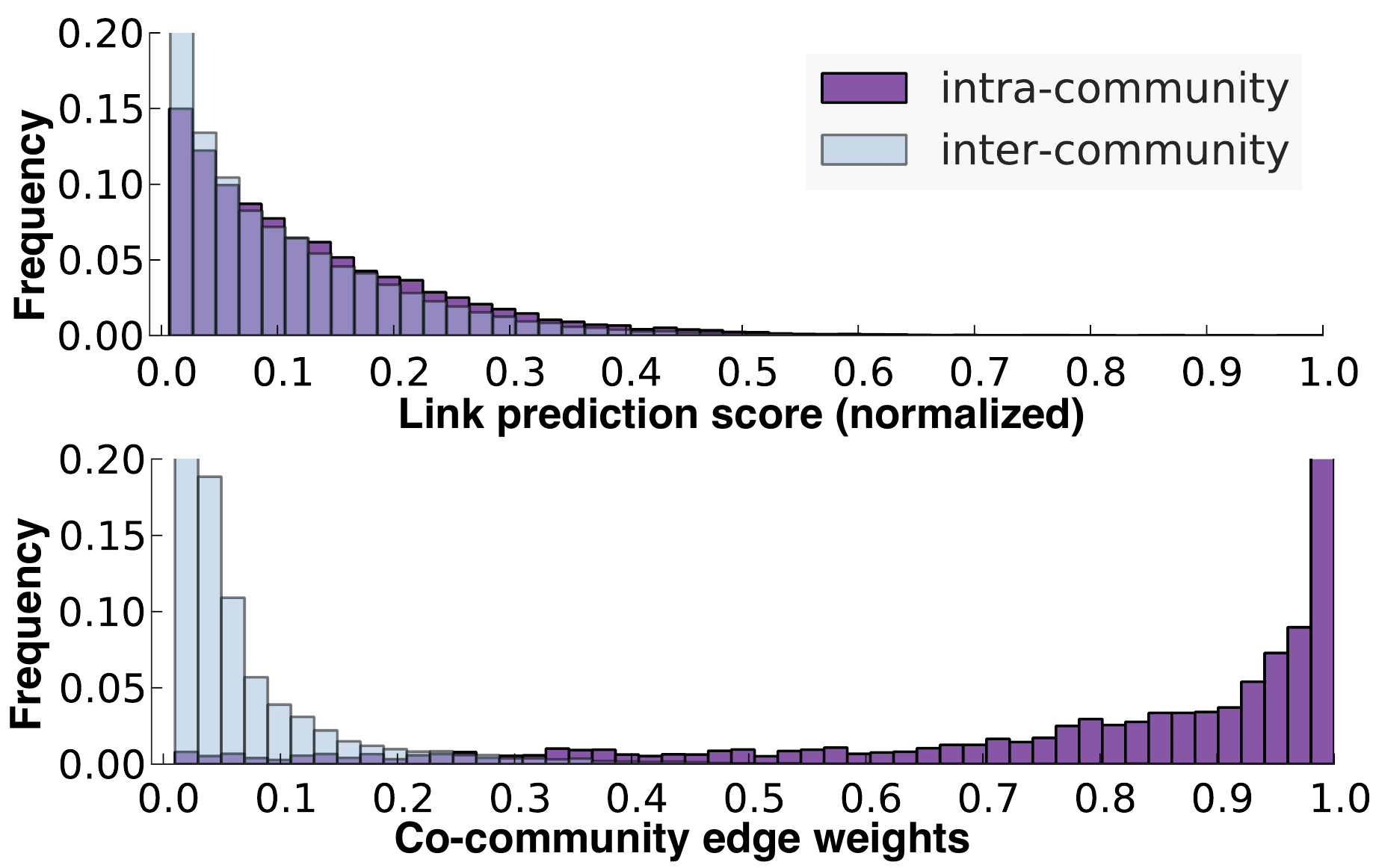}
  \caption{Histogram of edge weights on a benchmark graph with $\mu$=0.4 and 20\% of the edges removed: scores from AA link predictor (top) and weights of co-community network (bottom).}
\label{fig:edge_weight_histogram}
\end{figure}

An intuition for why this \naive~ algorithm does not work is illustrated in the top histogram of Figure~\ref{fig:edge_weight_histogram}. The plot shows the score distribution of both intra- and inter-community edges predicted by the $AA$ link predictor on a randomly generated benchmark network. The distributions of the intra-community edges substantially overlaps with the inter-community distribution, thereby making any choice of a threshold for adding links not helpful for community detection. In addition, as this plot shows, the top-$k$ edges only comprise of a small percentage of the total set of intra-community edges. By simply selecting from the top-k scoring edges, many of intra-community edges that are lower ranked will never be selected. As demonstrated in Figure~\ref{fig:link_prediction}, the choice of $k$ can have a significant impact on the quality of the edges, therefore selecting the right $k$ becomes a challenge when the complexity and sparsity of the network is unknown.

\section{Methods}
\label{sec:methods}

Our core observation is that link prediction in both high-$\delta$ and high-$\mu$ settings is brittle: it can carry information, but for a single prediction is likely to be wrong.  Therefore, we propose an improved method for applying link prediction to enhance community detection.

\floatname{algorithm}{Procedure}
\renewcommand{\algorithmicrequire}{\textbf{Input:}}
\renewcommand{\algorithmicensure}{\textbf{Output:}}
\algrenewcommand\Return{\State \algorithmicreturn{} }

\begin{algorithm}
\caption{\system Algorithm}\label{framework}

\begin{algorithmic}[1]
\Require A network $G = (V,E)$, link-predictor $\mathcal{L}$,
community detection algorithm $\mathcal{C}$, number of iterations $n$
\Ensure A partition $P^*$ of the vertices in $G$ 
\State $E_\text{missing} = \mathcal{L}(G)$ \Comment{score edges in $G$}
\State $\mathcal{D} = \textsc{Imputation}(E_\text{missing})$\Comment{create edge distribution}
\State $P = []$ \Comment{initialize list of partitions}

\For{$i\gets 1, n$}
\State $k \sim U(1,|E|)$
\State $\{e_1,e_2,\dots,e_k\} \sim \mathcal{D}$ \Comment{sample $k$ edges}
\State $G_i = (V,E \cup \{e_1,\dots,e_k\})$ \Comment{impute $G_i$}
\State $p_i = \mathcal{C}(G_i)$ \Comment{cluster $G_i$ }
\State $P = P \cup p_i$ 
\EndFor
\State $P^* = \textsc{AggregationFunction}(G,P)$
\Return $P^*$
\end{algorithmic}
\end{algorithm}

In order to mitigate the potential side effects of imperfect link prediction, we propose a sampling based algorithm that repeatedly applies link prediction to the input network. The \system\ pseudocode is shown in Algorithm~\ref{framework} and proceeds in four steps.  First, it uses a link prediction function to score missing edges and thereby construct a probability distribution over the set of missing edges (lines 2-3). The algorithm repeatedly samples a set of edges from this probability distribution, adding these sampled edges to the original network, and runs community detection on the enhanced network (lines 5-7). Each iteration produces a new set of communities which are added to the set of partitions (lines 8-9).  After the sample-detect-partition sequence is executed many times, we aggregate the overall set of observed partitions (line 11) to produce a final clustering.

\subsection{Network Imputation}
\label{sec:imputation}
The network imputation component of \system uses the input network and a link prediction algorithm to produce a probability distribution over the set of missing edges. The number of edges sampled during each iteration of the imputation procedure (lines 5-6) is a uniform random number between 1 and the size of the input network. We experimented with many values of $k$ and found this to work as well as when $k$ was fixed. 

We propose an imputation algorithm that constructs a distribution in which the probability of drawing an edge corresponds to its score produced by the link predictor. Missing edges that are scored higher by the link-predictor will have more probability mass than lower scoring edges. The probability function constructed from this process is:
\begin{equation}
 P(X=x) = \frac{\mathcal{L}(x)}{N} : x \in E_\text{missing},\ \
 N = \displaystyle{\sum_{x \in E_\text{missing}}\mathcal{L}(x)}
\end{equation}

Our imputation algorithm is more likely to pick higher scoring edges, which can result in a fairly accurate selection of intra-community edges as shown in section {\it Link-Enhanced Community Detection}. At the same time, even low scoring edges have probability mass, which is important since for some networks, intra-community edges can also be low scoring.

\begin{figure}[t]
  \centering
  \includegraphics[width=\columnwidth]{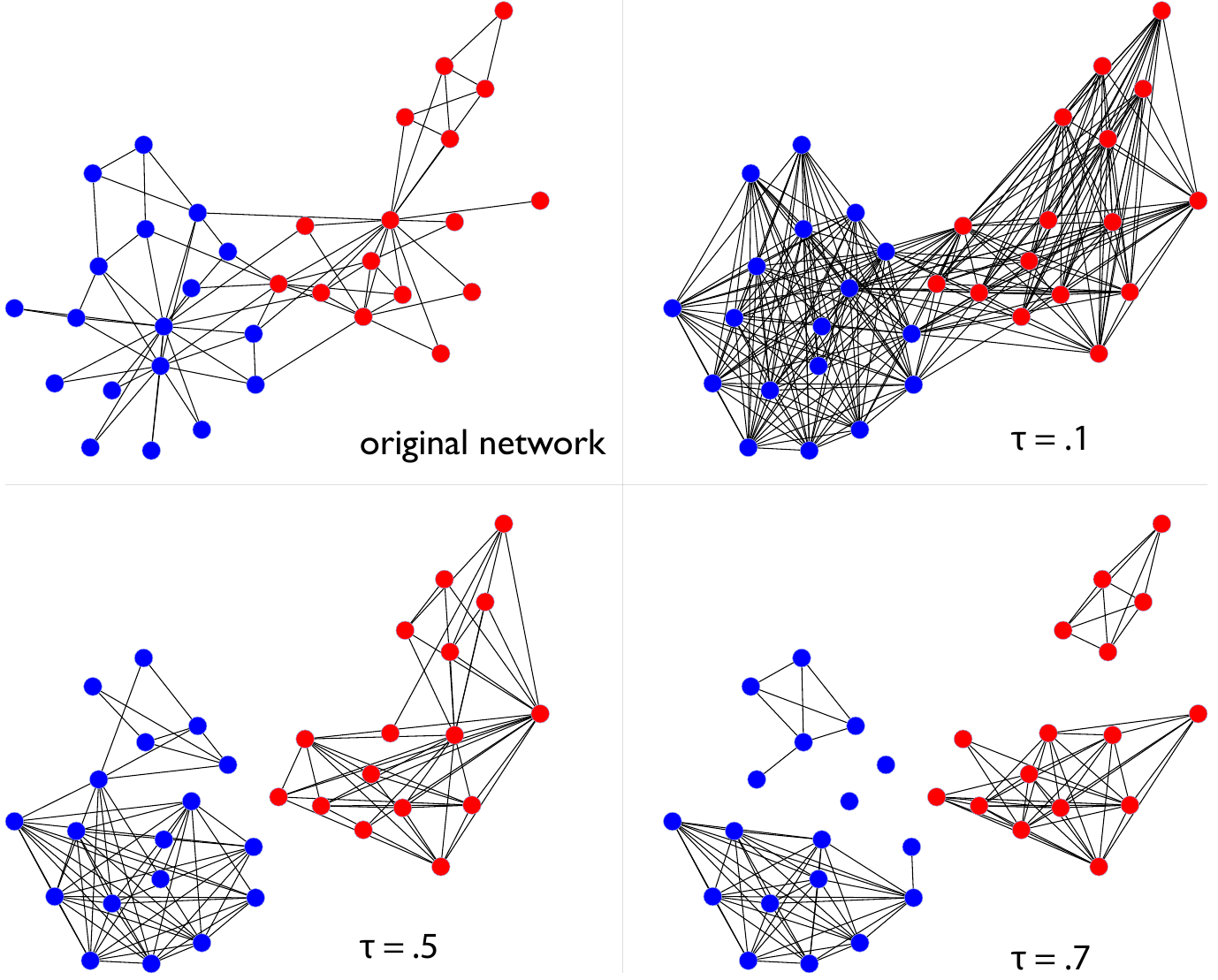}
\caption{Visualization of the co-community network for ``Zachary's karate club'' network. Each panel shows the network pruned at various thresholds $\tau$. }
\label{fig:karate_club}
\end{figure}

\subsection{Partition Aggregation}
\label{sec:partition_agg}

Having generated many possible ``images'' of our original graph via network imputation, we can apply community detection algorithms to each. Each execution of the algorithm produces a partition---possibly unique---based on the input graph. After generating many such partitions, we use {\em partition aggregation} to produce a final output. Previous ensemble clustering techniques~\cite{Dahlin_Ensemble,Lancichinetti_Concensus}, construct a $n\times n$ consensus matrix that represents the co-occurrence of nodes within the same community. The goal of such a data structure is to summarize the information produced by the various partitions. We propose a similar data structure, a co-community network $G_\text{cc}$, which consists of nodes from the input network and edges with weights that correspond to the normalized frequency of the number of times the two nodes appear in the same community.

The $G_\text{cc}$ graph is a transformation of the input network into one that represents the pairwise community relationships between nodes, rather than the functional relationships defined by the semantics of the input network $G$. $G_\text{cc}$ links nodes that appear in the same community, and weights them based on frequency or co-occurence (i.e., the edge between two nodes has a normalized weight equal to number of times the two nodes appear together in the same community over all partitions). Thus, $G_\text{cc}$ exhibits community structure representing communities that appeared frequently in the input partitions. As shown in the lower plot of Figure~\ref{fig:edge_weight_histogram}, there is a clear distinction between the intra-community and inter-community edge-weight distributions in $G_{cc}$. A simple mechanism for identifying a final ``partitioning'' is to remove all edges for which we have low confidence (i.e., inter-community edges) and studying the resulting connected-components (CC). We parametrize the pruning with a threshold $\tau$ and prune edges below that value. The semantics of the resulting graph is that all pairs of linked nodes have been seen in the same community at least $\tau$ percentage of times and consequently all nodes captured in a CC maintain this guarantee.

Figure~\ref{fig:karate_club} shows an example of a co-community network pruned at various thresholds. The network in this diagram is the famous {\textit Zachary's karate club}~\cite{Zachary_Network}, the colors of the nodes denote the ground-truth community assignments of each node. The original network is shown in the upper left quadrant, and the remaining quadrants show the co-community network pruned at different thresholds. As we can see in the upper right quadrant, if we threshold at a small value of $\tau$ we are likely to get nearly a fully connected network (and certainly one with only one large component). This is due to the fact that given enough iterations of the link prediction/community detection loop we are likely to find at least a few cases where nodes that would ordinarily fall into two communities are placed into the same one. At $\tau = 0.5$ we see the CC's reflect the community structure in the original network (the ``correct partition'').  As we increase the threshold the two true communities are further shattered into sub-communities, leaving some nodes completely isolated. One can interpret the connected components at these higher levels of $\tau$ as capturing the {\it core members} of the true communities: members who co-occur with each other a very high percentage of time and do not co-occur often with nodes outside of their community.

While $\tau$ may be set manually---appropriate for some applications when some level of confidence is desirable---there are other applications where we would prefer that this threshold be chosen automatically. As the last component of our framework we propose a way for selecting a $\tau$ and constructing a final partitioning given that chosen value. Since the edge weights in $G_{cc}$ correspond to the fraction of times two nodes appear in the same community, they are rational numbers. We can therefore enumerate all the possible values of $\tau$, $\{\frac{1}{n},\frac{2}{n},...,\frac{n}{n}\}$ on the interval $[0,1]$. At each value of $\tau$ we prune all edges with weights less than $\tau$ and compute the partition of $G_\text{cc}$ that corresponds to the connected-components. We then score this partition according to equation \ref{eq:stability_equation} and select the threshold and corresponding partition that maximizes this score. 

In previous work, Monti~\etal~\cite{Monti_Concensus} propose a formula for computing the ``consensus'' score of an individual cluster. For a given community $C_k$, that is of size $N_k$, their score sums the co-community weights and divides it by the maximum possible weight 

\begin{equation}
m_k = \frac{1}{\binom{N_k}{2}} \sum_{\substack{i,j \in C_k \\ i < j }}G_{cc}(i,j)
\end{equation}

We score a partition $p_\tau$ parametrized by a threshold $\tau$ by taking the weighted sum of the scores $m_k$ for each community in the partition. We use a weighted sum because the score contribution of each community should be commensurate with its size. 

\begin{equation}
\label{eq:stability_equation}
 S(p_\tau) = \frac{N_k}{N} \displaystyle{\sum_{k \in p_\tau}}m_k
\end{equation}

 If the final partition has any singleton nodes that don't belong to any community we connect each stray node to the community to which it has the highest mean edge weight to in the un-pruned co-community network. 

\section{Experiments}

We have conducted a series of experiments to test \system on the LFR benchmark networks, standard real-world networks (e.g karate club), and a set of ego networks from Facebook. First, we present a comparison of \system with different community detection methods. Subsequent experiments include an analysis of various parameter settings of \system.

\begin{figure}[]
  \centering
  \includegraphics[scale = 0.46]{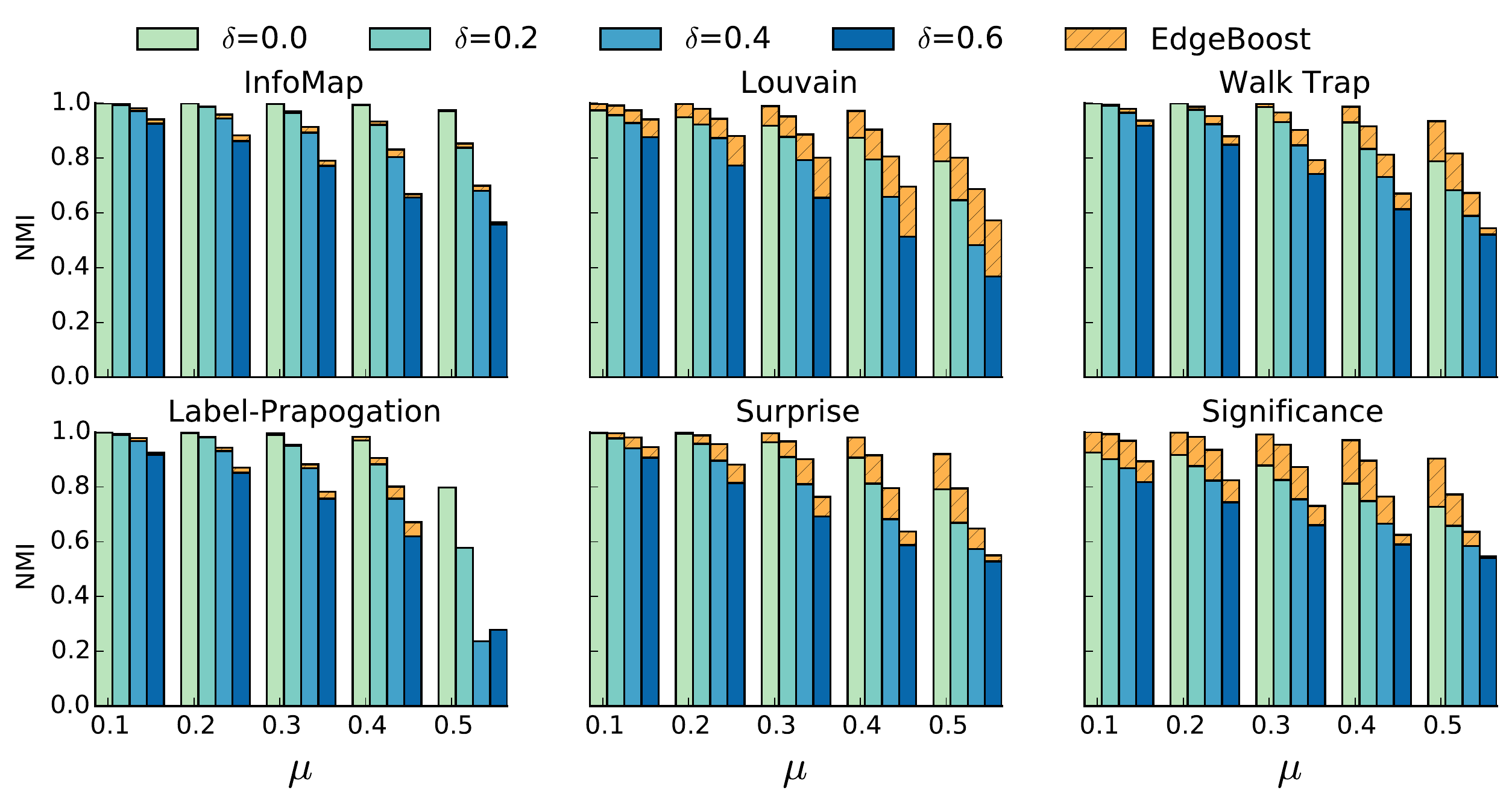}
  \caption{Performance of six popular community detection algorithms on the LFR benchmark networks. Dashed yellow bar shows the improvement of EdgeBoost over using the baseline community detection method.}
\label{fig:edgeboost_barchart_LFR}
\end{figure}

\begin{figure}[]
  \centering
  \includegraphics[scale = 0.46]{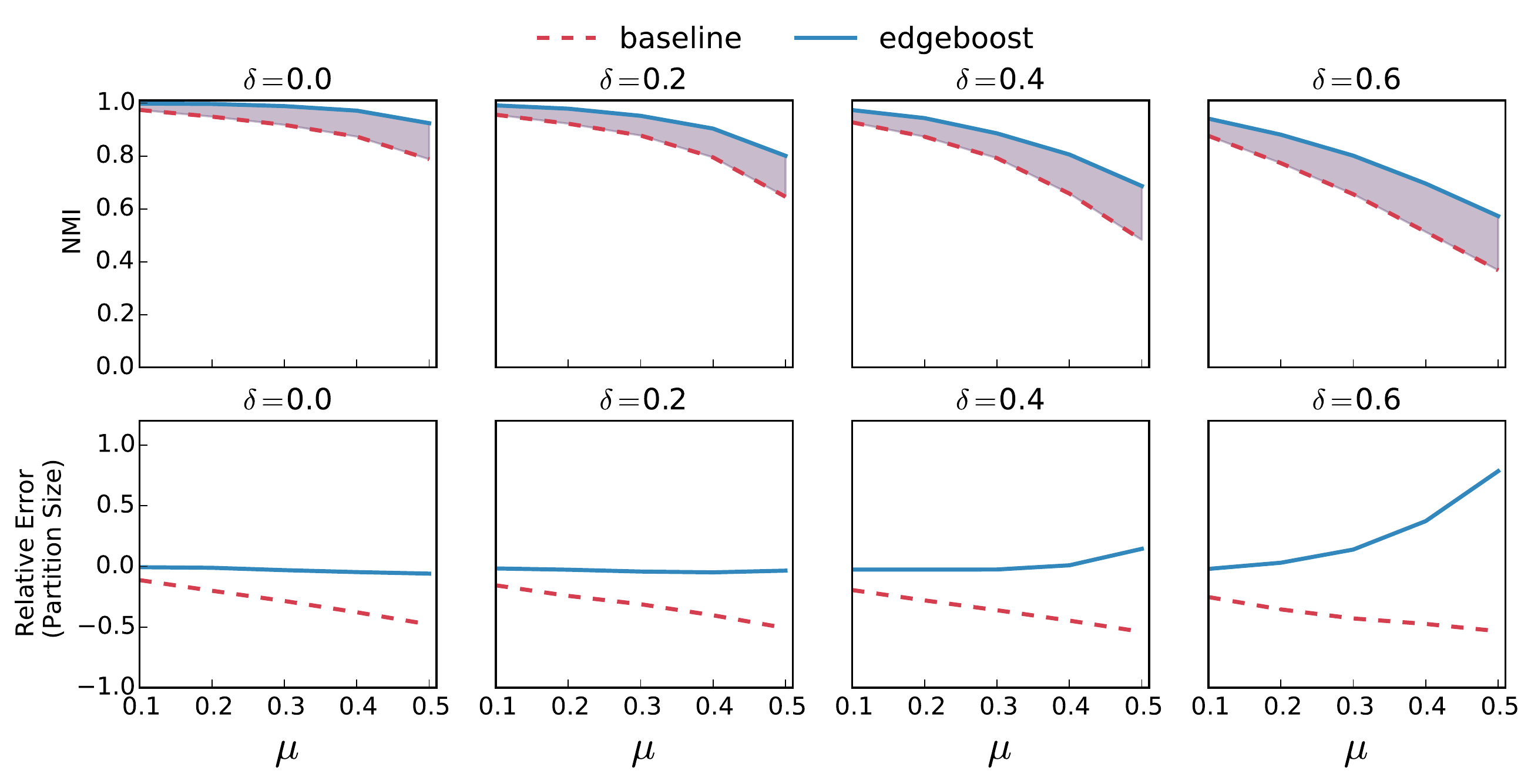}
  \caption{Performance of EdgeBoost (solid) and the baseline Louvain algorithm (dashed) on the LFR benchmark. The purple shaded region shows the improvement of edge boost for NMI. The bottom row of plots shows the relative error of the partition size. }
\label{fig:edgeboost_louvain_LFR}
\end{figure}

\subsection{Comparing \system with Different Community Detection Methods}

Similarly to the analysis in Section {\it Communities in Incomplete Networks}, we evaluate our methods against the LFR benchmark over various settings of the mixing ratio $\mu$ and the percentage of missing edges, $\delta$. In Figure~\ref{fig:edgeboost_barchart_LFR} we show the performance gain (striped yellow bars) of \system for six different community detection algorithms: InfoMap, Louvian, WalkTrap, Label-Propagation, Surprise, and Significance. The number of imputation iterations is fixed at 50 for both algorithms and the bars are generated by averaging over 50 randomly generated networks. While not shown in the figure, We tested all 3 link prediction algorithms and did not find a substantial difference. In agreement with our link prediction analysis in Section {\it Link Prediction for Enhancing Community Detection}, Jaccard slightly outperformed the other methods, so we chose Jaccard as the link prediction algorithm for \system. 

We can see from Figure~\ref{fig:edgeboost_barchart_LFR} that our method improves performance for almost all input community detection algorithms. One exception is that \system shows a decrease in performance for the Label-Propagation algorithms at a $\mu$ value of 0.5. As in other studies \cite{Lancichinetti_Concensus}, the Label-Propagation algorithm's performance becomes erratic at $\mu$ values of 0.5 or greater, most likely due to the fact that Label-Propagation assumes that a node's label should be chosen based on the labels of its neighbors. While \system is designed to work on stochastic algorithms, and variations of the input network, if an algorithms has too much variation, as is the case with Label-Propagation, it can lead to decreased performance.  While we can test the limits of community detection by setting high values of $\mu$, it has been shown that LFR networks with $\mu$ values of 0.5 and higher do not reflect the expected properties of real world networks~\cite{Orman_Comparison}. 

Figure~\ref{fig:edgeboost_louvain_LFR} shows the performance gain of \system on the Louvain algorithm in more detail. As our previous analysis showed, the baseline Louvain algorithm tends to detect bigger communities on average then in the planted partition. The bottom row shows that for moderate values of $\delta$, \system is able to recover the smaller communities in the planted partition. At very high values of $\delta$ ($>= 0.4$) a network may be so sparse then the perfect recovery of correct communities is most likely not possible. Even for these high $\delta$ values, \system still shows an improvement in NMI over the baseline method. 

\begin{figure}[t]
  \centering
  \includegraphics[width=\columnwidth]{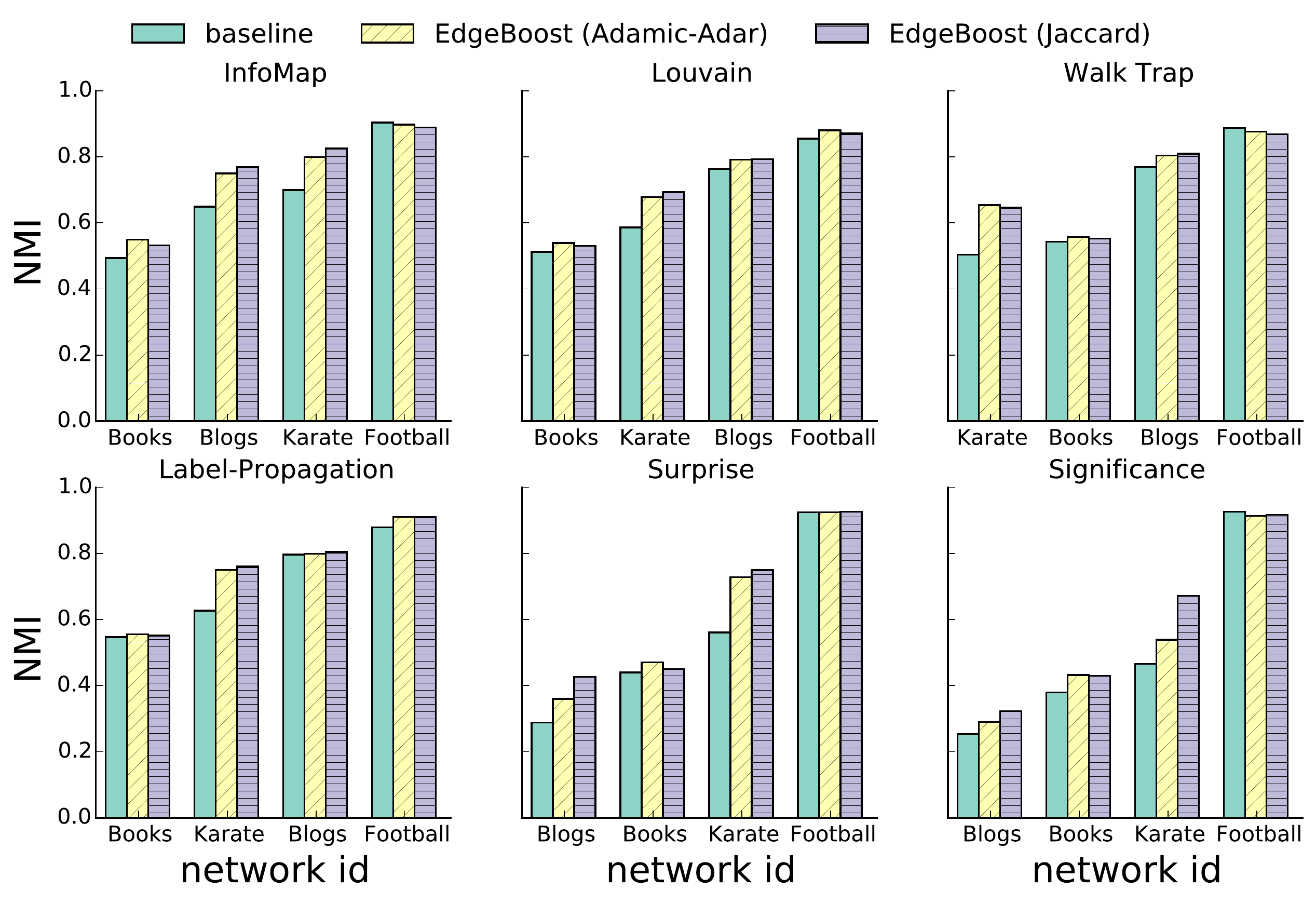}
  \caption{Comparison of \system on set of standard real network benchmarks community detection }
\label{fig:toy_networks}
\end{figure}

While the LFR benchmark captures certain properties, it is an imperfect model of real-world networks. To test \system on real network data, we also performed experiments on two real-world network data sets. The first data set consists of a suite of standard networks for benchmarking community detection. The data set includes: Zachary's {\it Karate Club} network (Karate)~\cite{Zachary_Network}, network of political books (Books) \footnote{This data set is not cited in the literature but can be found at http://www-personal.umich.edu/~mejn/netdata/}, blog network (Blogs)~\cite{Adamic_Political} and the American college football network (Football)~\cite{Girvan_Community}. All of these networks have a ground truth partition such that we can use NMI to evaluate the performance of community detection. Figure \ref{fig:toy_networks} shows the results of \system on each of the four networks with the same six input community detection algorithms used in the experiment above. In all but 3 of algorithm/network configurations, \system improves performance by an average of 14\%. On the Football network, \system does worse with the InfoMap, Label-Propagation, and WalkTrap algorithms, but decreases performance by only an average of 1.6\%. Overall, this dataset gives some assurance that \system can improve performance on real networks.

We also test \system on a data set of Facebook ego-networks~\cite{McAuley_Learning} that capture all neighbors (and their connections) centered on a particular user. The data set described in the original paper by McAuley~\etal, consists of networks from three major social networks: Facebook, Google+ and Twitter. The Facebook data set is likely the highest quality of the three, it contains ground-truth which was obtained from a user survey that had the ego users for each network provide community labels \footnote{The Facebook dataset can be found at http://snap.stanford.edu/data/egonets-Facebook.html}. The ground truth for the ego-networks from Twitter and Google+ is lower quality since it was obtained by crawling the publicly available lists created by the ego user. As such, for many of the networks, the ground truth consisted of only a small fraction of nodes in the network and for many networks the ground truth consisted of lists with very few members. Since the target of this paper is non-overlapping and complete clustering, we chose to not use the Twitter and Google+ networks due to the sparsity of their ground-truth. The Facebook networks have complete ground-truth labeling, so we used those for evaluation. Despite the Facebook networks being the highest quality of the three datasets, it still contained ground-truth communities of 1-2 users. We pre-processed each network by removing all ground-truth communities with fewer then three nodes.

\begin{figure}[t]
  \centering
  \includegraphics[width=\columnwidth]{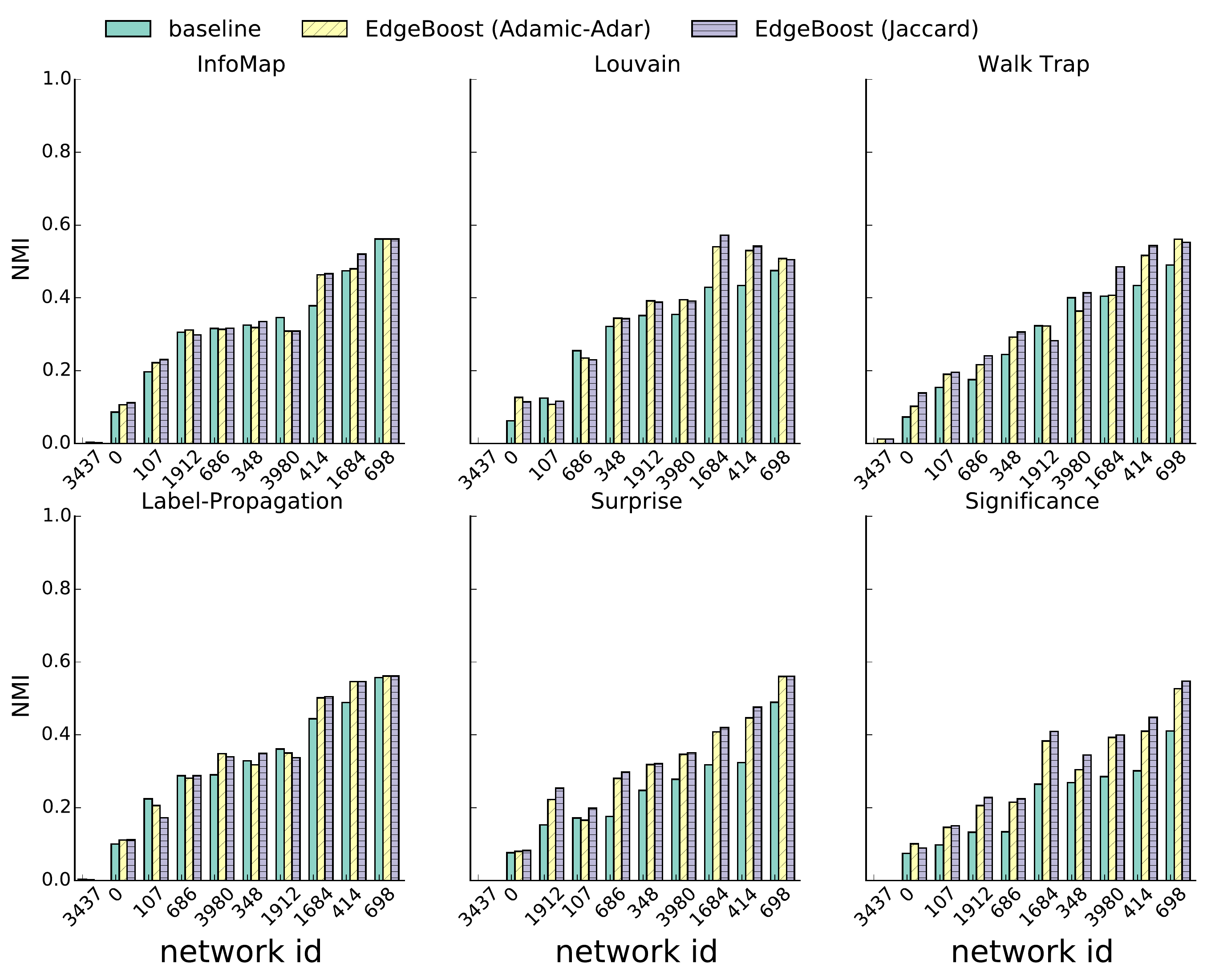}
  \caption{Comparison of \system on ego-networks from Facebook}
\label{fig:ego_nets_results}
\end{figure}
    As the ground-truth for the ego-networks in all data sets can contain overlapping communities, we cannot directly use the standard version of NMI for evaluation. Lancichinetti~\etal~\cite{Lancichinetti_Detecting} propose an extension of NMI that can compare overlapping communities which we use to test \system on the overlapping ground-truth data. Figure~\ref{fig:ego_nets_results} shows the results of using \system with the same six community detection algorithms used in the LFR experiments. The solid bars represent the performance of the baseline community detection algorithm without \system. The diagonal and horizontal striped bars shows the results from \system paired with the Adamic-Adar and  Jaccard respectively. We set the number of iterations for \system at 50. each bar was generated by averaging the NMI score over 100 runs of the baseline and \system paired with the Jaccard and Adamic-Adar link predictors. \system shows an improvement on most networks for each of the six community detection algorithms; this result is consistent with our experiments on the LFR benchmark. On the LFR benchmark networks \system paired with Jaccard link prediction was consistently better then the other link prediction methods but this is not consistently the case on the Facebook networks. Jaccard outperforms the Adamic-Adar most of the time, but there are some cases when the opposite is true. While \system shows improvement for most combinations of algorithms and network, there are some instances when the performance of \system is lower then baseline. Overall, in 52 of the 60 total configurations \system improves performance by an average of 21\%. In the rare configurations (8 out of 60) when \system performs worse then baseline, \system performs only 5\% worse on average.  

\subsection{Varying the Parameters of \system}

\begin{figure}[h]
  \centering
  \begin{tabular}{c}
    \subfloat[\system with Louvain]{\label{fig:num_iterations_louvain}
      \includegraphics[width = \textwidth]{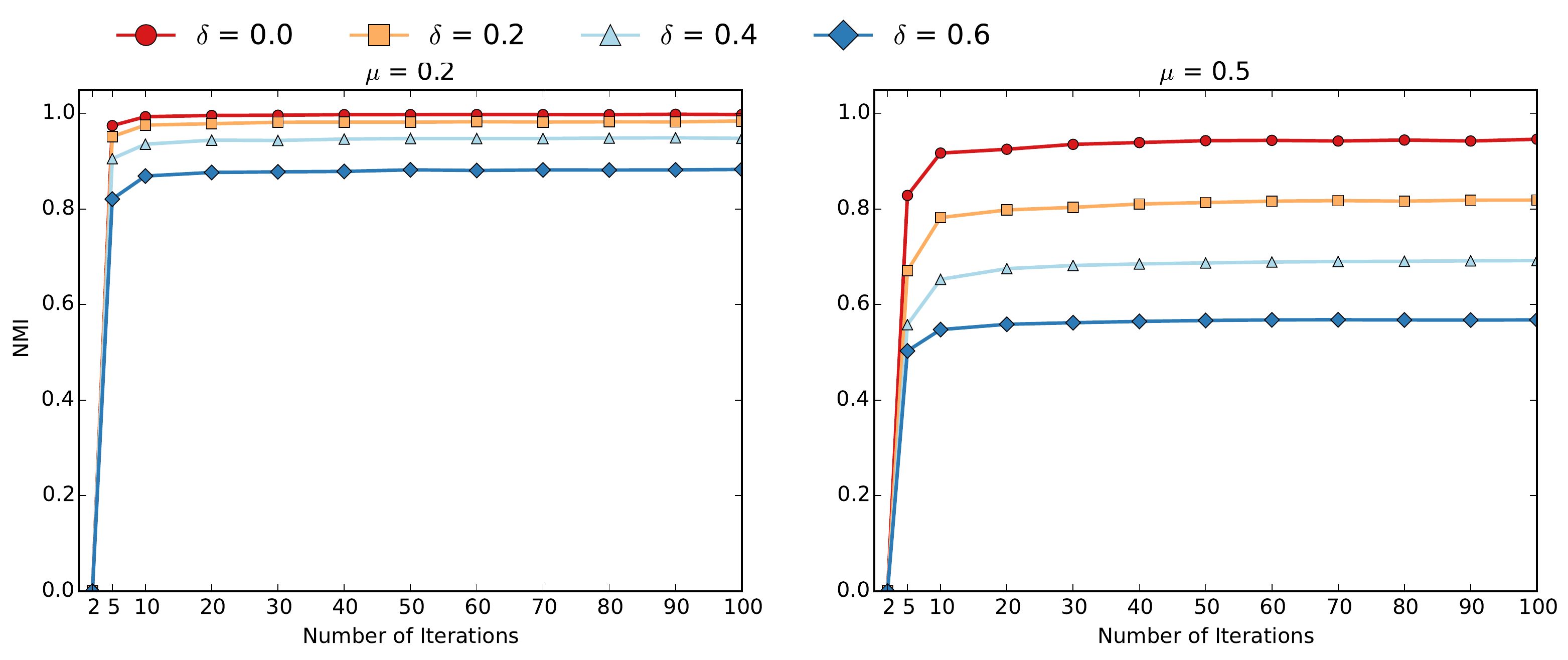}} \\
    \subfloat[\system with InfoMap]{\label{fig:num_iterations_infomap}
      \includegraphics[width = \textwidth]{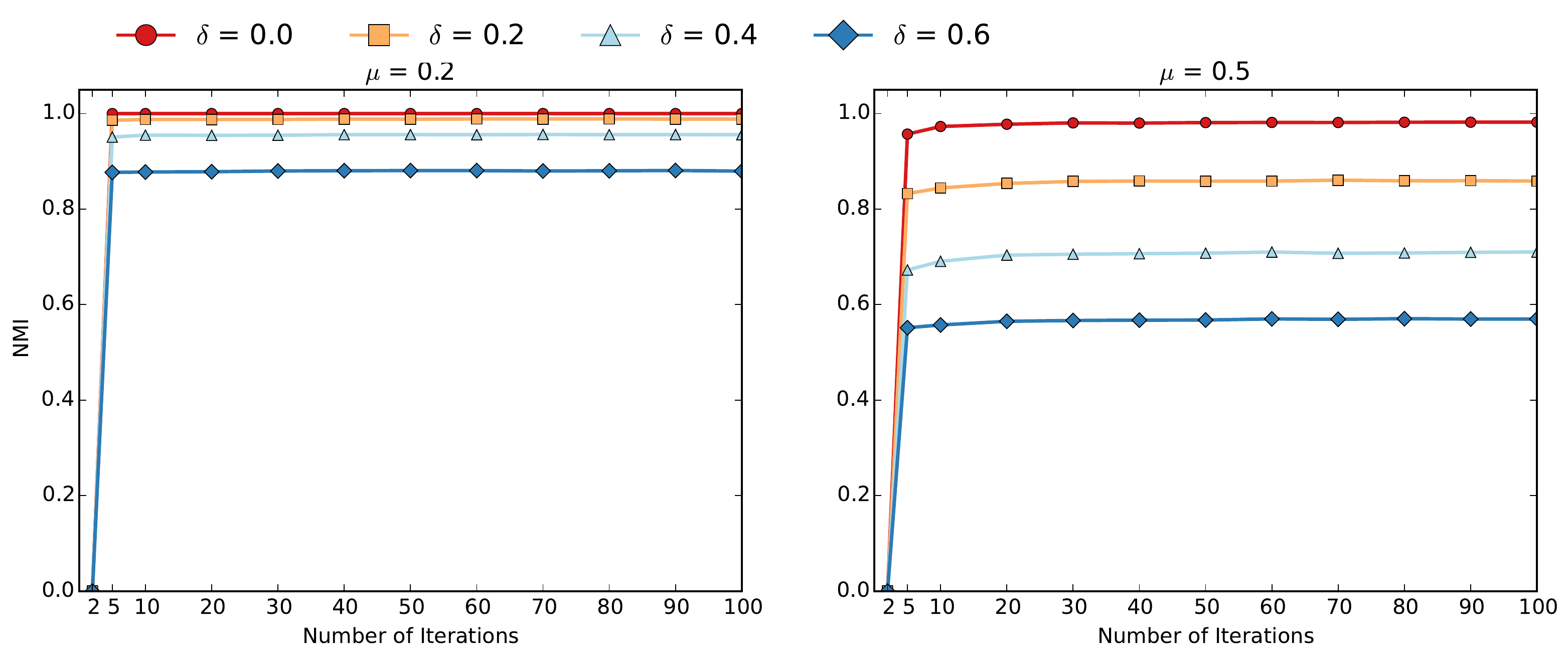}} 
  \end{tabular}
  \caption{Varying the number of iterations ({\it NumIterations}) for with $\mu = 0.2$ (left) and $\mu = 0.5$ (right) over $\delta$ values ranging from 0.0 to 0.6.}
  \label{fig:num_iterations}
\end{figure}



In addition to comparing \system using different community detection algorithms we also analyzed how the performance varies with respect to different parameter settings. For these experiments, the curves were generated by averaging over 50 networks generated via the LFR benchmark. Figures~\ref{fig:num_iterations_louvain} and \ref{fig:num_iterations_infomap} show the convergence of the Louvain and InfoMap algorithms, as a function of the ``number of community detection iteratations'' ({\it NumIterations}). Most of the performance gain from \system can be had with {\it NumIterations} set to 10, and setting the number of iterations beyond 50 doesn't give much benefit. The convergence of \system is qualitatively similar for low and high values of $\mu$ and the entire range of $\delta$ values.

\begin{figure}[t]
  \centering
  \begin{tabular}{c}
    \subfloat[\system with Louvain]{\label{fig:threshold_analysis_louvain}
      \includegraphics[width = \textwidth]{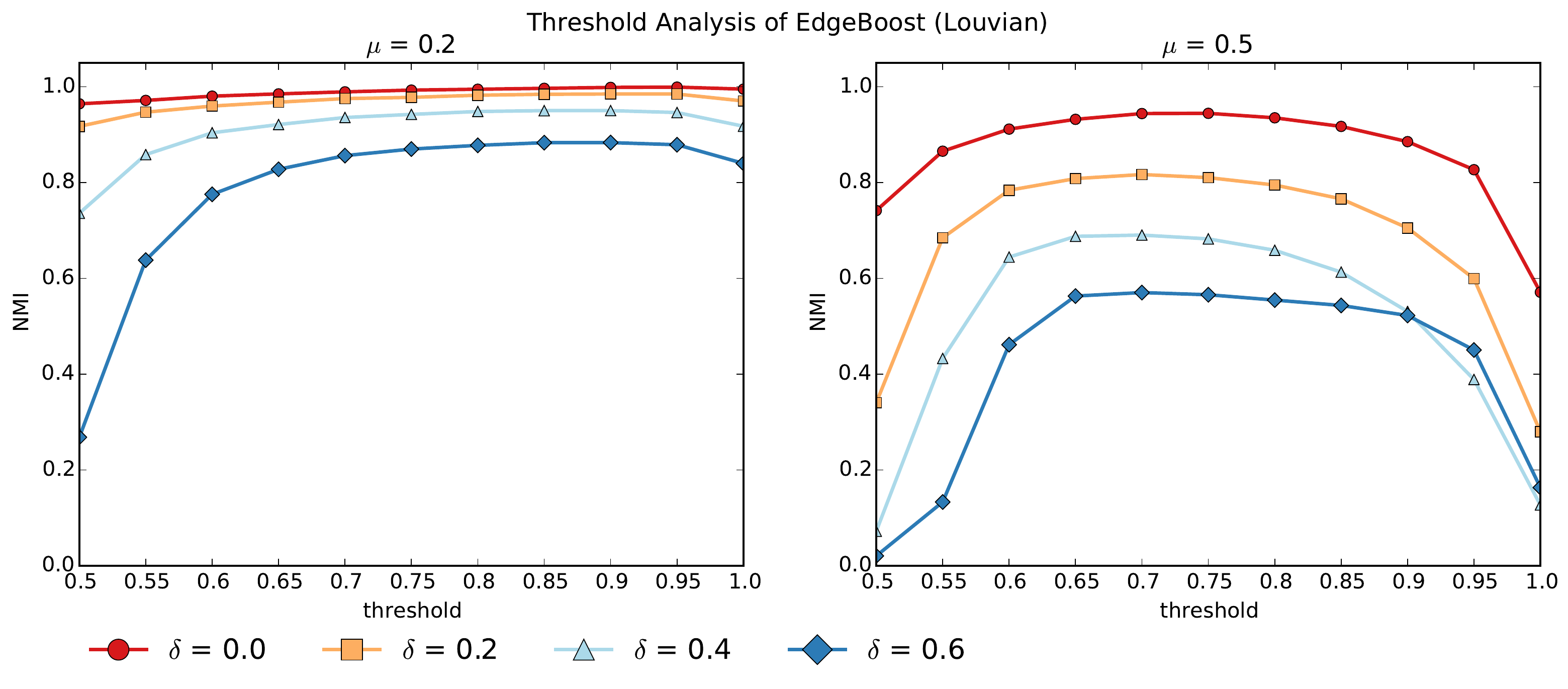}} \\
    \subfloat[\system with InfoMap]{\label{fig:threshold_analysis_infomap}
      \includegraphics[width = \textwidth]{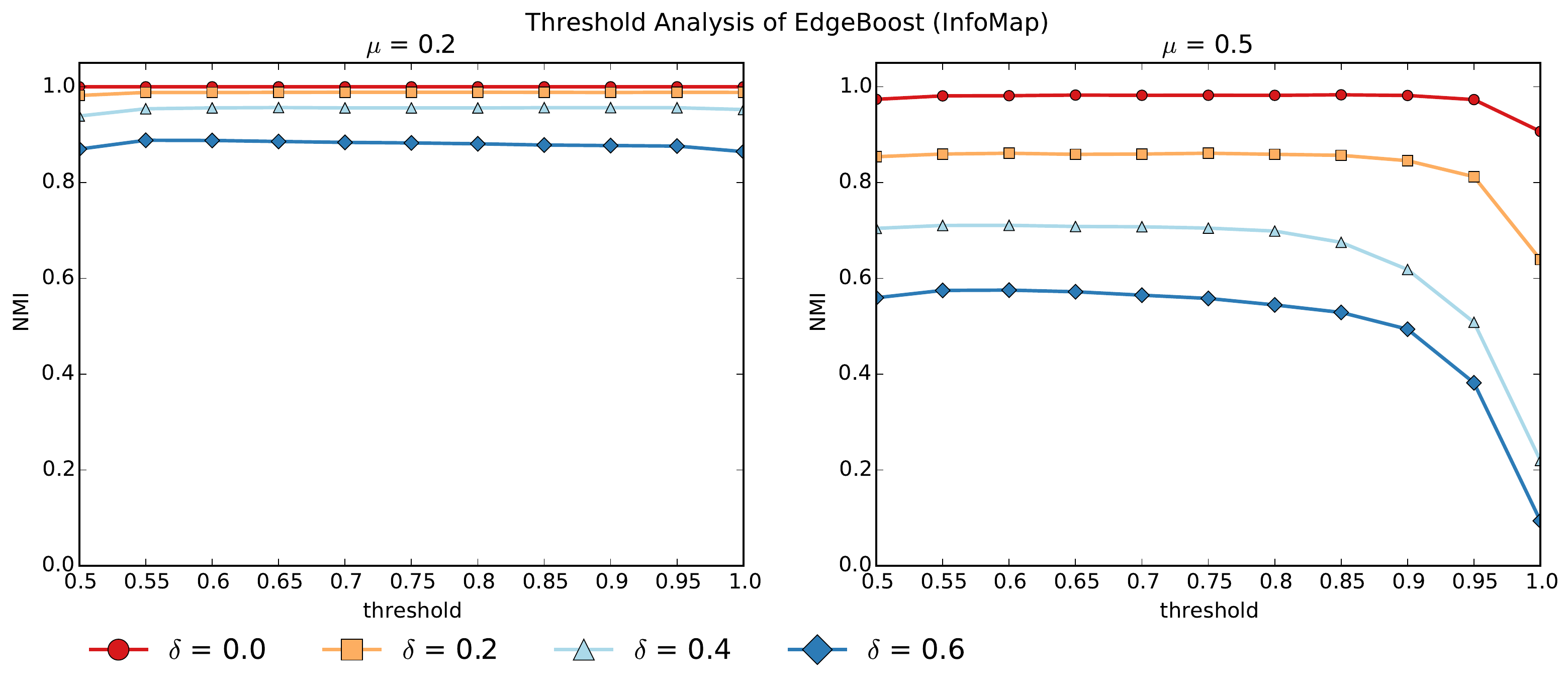}} 
  \end{tabular}
  \caption{Varying the co-community threshold ($\tau$) for \system with $\mu = 0.2$ (left) and $\mu = 0.5$ (right) over $\delta$ values ranging from 0.0 to 0.6.}
  \label{fig:threshold_analysis}
\end{figure}

In Section {\it Partition Aggregation} we propose a method for automatically selecting the co-community threshold $\tau$, which we have used for all of the previous experiments. Since the selection of $\tau$ is the most computationally expensive parts of the entire \system pipeline, we present an analysis of how \system performs with a manual selection of $\tau$. Figures~\ref{fig:threshold_analysis_louvain} and \ref{fig:threshold_analysis_infomap} show how \system performs by varying the selection of $\tau$ for \system paired with Louvain and InfoMap respectively. For both algorithms, \system can achieve good performance for values of $\tau$ in the range 0.6-0.9, indicating that manual $\tau$ selection can be an effective way to save computational resources and still boost performance over baseline. For higher values of $\mu$, the performance of \system is more dependent on $\tau$, especially for the Louvain algorithm. Since the Louvain algorithms performs less reliably for higher $\mu$ values, the co-community network has noisier edge weights, therefore making the selection of $\tau$ more critical to achieving good performance.

\begin{figure}[t]
  \centering
  \includegraphics[scale=0.4]{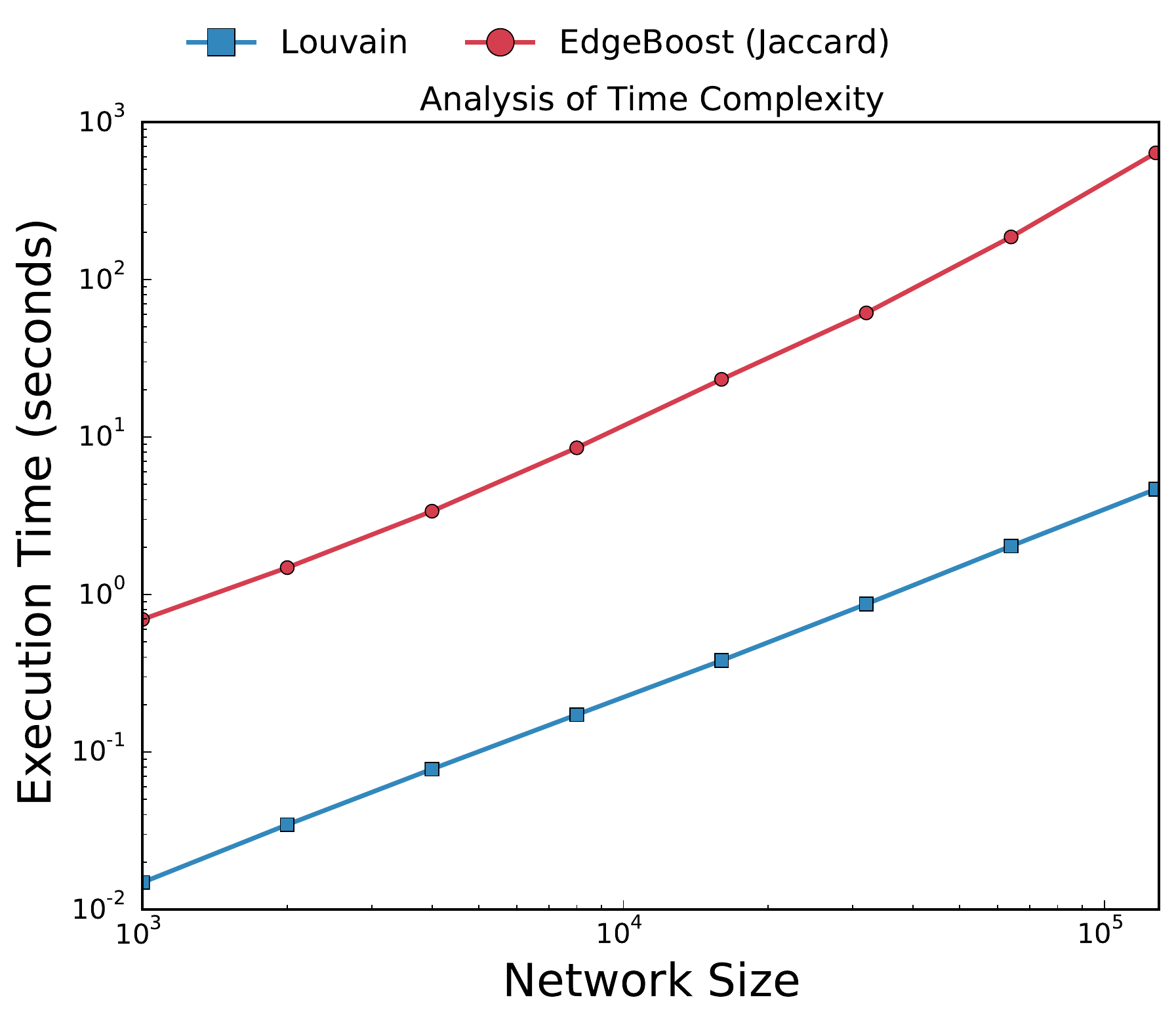}
  \caption{Analysis of execution time}
\label{fig:time_analysis}
\end{figure}

\subsection{Runtime Analysis}

The most computationally expensive module of \system is the aggregation algorithm which requires the computation of connected components at various thresholds. The complexity of computing connected components is worst case $O(|E|)$, where $|E|$ is the number of edges in the network. The aggregation module computes the connected components on the co-community network, which can be much denser then the input network. In theory it is possible for the co-community network to have $O(n^2)$ number of edges, therefore making the aggregation module computationally expensive. In order to show that \system scales well when increasing to large networks we ran it on LFR networks of various sizes, ranging from 1000 to 128000 nodes. Figure \ref{fig:time_analysis} shows the run time of running \system with Louvain and Jaccard link prediction with the number of iterations set to 10. While \system does have a significant time overhead over Louvain, it still scales in the same manner as Louvain.


\section{Discussion and Future Work}

\label{sec:discussion}

Our modified LFR benchmark used random edge deletion to model missing edges in networks. While we chose this model because we think that it is the most generally applicable, there are other possibilities for modeling missing edges. One area of future research is to see how community detection algorithms are affected using different edge deletion strategies. Further experiments are necessary to determine if our techniques withstand biased edge removal, but we believe that repeated link prediction will nonetheless boost performance. Further, if missing intra-community edges could be modeled more accurately, development of better link prediction algorithms may be possible.

In order to increase the quality of community detection, \system trades off time and space efficiency. The construction of the co-community network can be memory consuming because it is likely to be much denser then the input network. In addition, \system  requires many runs of a sometimes costly community detection algorithm. While \system can scale to reasonably large networks (see section {\it Runtime Analysis}), we acknowledge these trade-offs and emphasize that \system is not designed for million node networks. Instead it was designed for use on small and medium networks (i.e ego-networks, citation networks), in which data sparsity problems are common, and communities reflect meaningful structures in the data.

While we have shown the efficacy of \system in computing better partitions, it is possible that the approach can also improve other types of community analysis. Given different thresholds for which we can prune the co-community network (see {\it Partition Aggregation} ) and the corresponding set of connected components, we can obtain a set of communities with a specified confidence. Some applications may not require a complete partitioning of nodes and may even be better suited with an incomplete partition which has higher quality communities. In future work we would also like to see how \system can be used in the detection of overlapping and/or hierarchical communities. This extension would require a different aggregation function as our current method is only capable of creating strict partitions, via computing connected-components.

The link-predictors tested in this paper are all based on shared neighbors, and therefore are only capable of inferring missing connections between nodes that are at maximum 2 hops from each other. One issue with predicting links that are further apart is the computational complexity, since most of the metrics that are not neighborhood based are based off the number of shortest paths between pairs of nodes. While not presented in this paper we experimented with the local path index proposed by~\cite{Lu_Similarity}, which predicts links between nodes that are as far as 3 hops from each other, but did not see any noticeable improvement. Other link predictors that we did not explore are those that utilize node attributes (\eg, school and city) and/or link structure to score missing edges.  Since most of the methods in disjoint community detection do not account for node attributes, in the future \system\ could be a robust way to integrate node attributes into existing algorithms. 


\section{Conclusions}

Networks inferred or collected from real data are often susceptible to missing edges. We have shown that as the percentage of missing edges in a network grows, the quality of community detection decreases substantially. To counter this, we proposed \system as a framework to improve community detection on incomplete networks. \system is capable of improving all the community detection algorithms we tested with its novel application of repetitive link prediction, on real ego-networks from Facebook, Twitter and Google+. \system is an easy-to-implement meta-algorithm that can be used to improve any user-specified community detection algorithm and we anticipate that it will be useful in many applications. 
\section{Acknowledgements}
This work was supported by National Science Foundation grant IGERT-0903629.



\bibliographystyle{abbrv}
\fontsize{9.5pt}{10.5pt} \selectfont
\bibliography{ref}

\end{document}